\documentclass[superscriptaddress,showpacs,twocolumn,aps,pre,10pt]{revtex4-1}
\usepackage{graphicx}
\usepackage[caption=false]{subfig}
\usepackage{amsmath}
\begin{document}

\title{Scaling of cluster growth for coagulating active particles}
\author{Peet Cremer}
\email{pcremer@thphy.uni-duesseldorf.de}
\affiliation{Institut f{\"u}r Theoretische Physik II: Weiche Materie, 
Heinrich-Heine-Universit{\"a}t D{\"u}sseldorf, D-40225 D{\"u}sseldorf, Germany}
\author{Hartmut L{\"o}wen}
\affiliation{Institut f{\"u}r Theoretische Physik II: Weiche Materie, 
Heinrich-Heine-Universit{\"a}t D{\"u}sseldorf, D-40225 D{\"u}sseldorf, Germany}
\date{\today}

\begin{abstract}
Cluster growth in a coagulating system of active particles (such as 
microswimmers in a solvent) is studied by theory and simulation. In contrast 
to passive systems, the net velocity of a cluster can have various scalings 
dependent on the propulsion mechanism and alignment of individual particles. 
Additionally, the persistence length of the cluster trajectory typically 
increases with size. As a consequence, a growing cluster collects neighbouring 
particles in a very efficient way and thus amplifies its growth further. This 
results in unusual large growth exponents for the  scaling of the cluster size 
with time and, for certain conditions, even leads to ``explosive'' cluster 
growth where the cluster becomes macroscopic in a finite amount of time.
\end{abstract}

\maketitle
\section{Introduction}
\label{sec.intro}
Phase separation of a homogeneous state into two distinct bulk phases is not 
only relevant for many technological processes, but also constitutes a 
classical problem of nonequilibrium statistical mechanics 
\cite{Binder1976_AdvPhys,Onuki2002_book,Tanaka2000_JPhysCondensMatt,
Lowen1997_PhysicaA}. For ordinary fluids and solids, the separation process is 
usually triggered by an initial fluctuation from which a critical nucleus 
arises. This initial cluster grows and ripens according to different scaling 
laws \cite{Krug1997_AdvPhys}. Typically, the extension of the cluster 
increases with a power law $R(t) \sim t^\alpha$ of time, where the exponent 
$\alpha$ depends on the growth process and the dimensionality $d$ of the 
system. For ordinary (passive) systems, $\alpha$ varies in the range between 
$1/3$ and $1$ \cite{Bray2002_AdvPhys}. More recently, both for mesoscopic 
colloidal suspensions \cite{Aarts2005_NewJPhys} and for complex plasmas 
\cite{Wysocki2010_PRL}, the phase separation process has been studied by 
observing the individual particle trajectories, giving insight into the 
microscopic (i.e. particle--resolved) mechanisms of the separation process 
\cite{Ivlev2012_book}.

While the physics of the phase separation processes is by now well--studied 
and understood for inert, passive particles, there is recent work 
demonstrating that similar separation and clustering processes occur for an 
ensemble of microswimmers. The latter can be regarded as active particles in a 
solvent (experiencing a Stokes drag) with an internal propulsion mechanism. In 
fact, there are widely different realizations of such active particles, 
ranging from swimming bacteria to artificial self--propelled colloidal 
particles 
\cite{Vicsek2012_PhysRep,Romanczuk2012_EPJST,Cates2012_RepProgPhys}. 

Basically, two different separation processes in active systems occur. First, 
clustering can be purely motility induced \cite{Cates2013_EPL}, such that it 
vanishes if the self--propulsion is removed as recently demonstrated 
\cite{Palacci2013_Science,Peruani2006_PRE,Ishikawa2008_PRL,Wensink2008_PRE,
Redner2013_PRL,Buttinoni2013_PRL,Bialke2013_EPL,Fily2012_PRL,Yang2008_PRE}. 
The simplest variant is a swarm of self--propelled particles resulting in an 
overall moving cluster. Second, there is already phase separation in the 
unpropelled, passive system which is then altered due to the drive. This was 
considered e.g. for a self--propelled Lennard--Jones system with attractive 
particle interactions \cite{Theurkauff2012_PRL,Gregoire2003_PhysicaD}. 
Attraction can hardly be avoided in metal--capped colloidal swimmers due to 
the mutual van--der--Waals forces \cite{Theurkauff2012_PRL}. However, for 
active particles, the dynamical evolution of cluster growth, as characterised 
by a nontrivial growth exponent $\alpha$, has only rarely been considered 
apart from very recent studies 
\cite{Redner2013_PRL,Redner2013_PRE,Wysocki2013_Arxiv,Mehes2012_PLoSONE,
Peruani2013_NewJPhys}.

In this paper, phase separation is investigated in a situation, where active 
particles irreversibly coagulate with each other on contact, resulting in a 
compact aggregate. Irreversible coagulation is well understood for passive 
particles 
\cite{Meakin1984_PRA,Meakin1985_PRA,Meakin1985_PRB,Kolb1983_PRL,Meakin1986_PRA}
and the scaling of cluster size with time has been studied as well 
\cite{Trizac1996_JStatPhys,Carnevale1990_PRL,Kolb1984_PRL}. We show here that 
activity of particles enables qualitatively different and novel cluster growth 
behaviour. Due to the self--propulsion, clusters perform a persistent random 
walk \cite{Soto2013_PRE} in contrast to the typical diffusive motion of 
passive particles. This allows a cluster of active particles to effectively 
``sweep up'' smaller clusters, which self--accelerates and amplifies cluster 
growth considerably further. Our theoretical analysis and computer simulation 
show that the cluster growth scaling exponent $\alpha$ can not only be 
considerably larger for active particles than the known values for passive 
particles, but that there is even a scenario of ``explosive'' cluster growth. 
We refer to the term ``explosion'' if the cluster reaches macroscopic size in 
a finite amount of time. Such explosive behaviour was found earlier in the 
context of gelation kinetics (see, e.g. \cite{Ben-Naim2003_PRE,
Herrero2000_PhysicaD,Singh1996_JPhysAMathGen,Dongen1988_JStatPhys}) and in 
phase separation in external fields, like gravity \cite{Falkovich2002_Nature}.

In detail, the growth exponent $\alpha$ depends on the scaling of the total 
propulsion force of a cluster with its size, the persistence of the cluster 
trajectory and the dimension $d$. We present several cases for the scaling of 
the total propulsion force of a cluster, which is determined by the type of 
swimmer, the fraction of particles contributing to the propulsion and the 
alignment of particles. If the cluster is driven by aligned surface particles 
only, in $d = 3$ dimensions we find up to exponential growth. Uncorrelated 
contribution of all particles leads to algebraic growth with up to 
$\alpha = 2$. Finally, if all particles in the cluster propel the cluster in 
the same direction ``explosive'' growth becomes possible. These results apply 
for the case that clusters possess a compact structure. Additionally, we also 
consider the case where the growing cluster is fractal and discuss briefly 
the scaling implications on the growth laws. All our predictions are verifiable 
in experiments for self--propelled particles with very strong 
van--der--Waals attraction, e.g. as prepared in Ref. 
\cite{Theurkauff2012_PRL}, phoretic attraction \cite{Palacci2013_Science} or 
dipolar interaction \cite{Baraban2013_ACSNano}.
\section{Scaling theory}
\label{sec.model}
We perform our scaling theory in a general $d$--dimensional space ($d=2,3$)
and assume that self--propelled particles irreversibly coagulate and form 
clusters with $N$ member particles and radius $R_N$ such that 
$R_N \propto N^{1/d}$. In the following, we refer to $N$ as the cluster size. 
The cluster formation process is described in a simplified way insofar as we 
consider compact clusters only and distinguish between different extreme 
cases. Once the particles contribute to the cluster, they stay fixed and their 
direction of self--propulsion (or orientation) is frozen. One may therefore 
distinguish two basic cases, one, where all orientations of cluster particles 
are completely uncorrelated and another where all directions are perfectly 
aligned. The first case occurs if the orientational reordering is frozen--in 
during coagulation (as realised for rough spheres) while the latter case 
arises if there is a considerable alignment interaction during the coagulation 
process (as realised for example for rod--like artificial swimmers or 
bacteria). The next basic distinction concerns the particles which really 
contribute to the overall self--propulsion of the cluster. Here we also 
discuss two extreme cases: either all cluster particles contribute in the same 
way or only particles at the cluster boundary contribute. The first case is 
realised for two--dimensional catalytic swimmers on a substrate which are 
embedded in a bulk liquid such that there is enough fuel all over the cluster. 
It also occurs for coagulation of passive colloidal particles in gravity 
\cite{Falkovich2002_Nature}. The second case of cluster surface activity is 
realised for three--dimensional catalytic swimmers where a fuel--depletion 
zone is created inside the cluster which reduces the push of inside--particles 
\cite{Paxton2004_JAmChemSoc,Ebbens2012_PRE,Golestanian2007_NewJPhys}. 
Moreover, catalytic swimmers move along the gradient of the chemical which 
also results in surface activity of the growing cluster 
\cite{Paxton2004_JAmChemSoc,Ebbens2012_PRE,Golestanian2007_NewJPhys}. 
Surface cluster activity also occurs due to hydrodynamics for pushers and 
pullers. When swimming in a tight formation, the propulsion of particles can 
be cancelled by the flow created by the swimmers behind them. Consequently, 
only the particles in the rear of the cluster contribute to the total 
propulsion force \cite{Cisneros2007_ExpFluids}, which again scales with the 
surface of the cluster.  

A single swimmer is propagated formally by an internal force
\footnote{It has been argued that real swimmers are force free and therefore 
do not directly feel a Stokes drag. However, the equations of motion can be 
expressed by a formal force which is proportional to the Stokes drag, see 
\cite{Golestanian2008_EurPhysJE}. Therefore, all scaling relations are unaffected by 
assuming an effective Stokes drag propulsion force.}
which is compensated by the Stokes drag at low 
Reynolds number resulting in a constant propagation velocity $v^{(0)}$. All 
these individual forces  $\mathbf{F}^{(i)}$ ($i=1,\dots,N$) add up to give the 
total force $\mathbf{F}_N$ acting on the cluster of size $N$ and putting it 
into motion with a velocity $v_N$. This force $\mathbf{F}_N$ is balanced by 
the Stokes drag acting on the cluster which scales in both $d = 3$ and $d = 2$ 
\cite{Kim1991_book} as $F_N \propto v_N R_N$. This after all yields different 
scalings for $F_N \propto N^{\beta}$ with a nontrivial exponent $\beta$ such 
that 
\begin{equation}
  v_N \sim F_N/R_N \sim N^{\beta - 1/d} .
  \label{eq.ClusterVelocity}
\end{equation}

\begin{figure}[htb]
  \centering
  \subfloat[][$\beta = 1 + \gamma$]{
    \centering
    \includegraphics[width = 0.4\columnwidth]{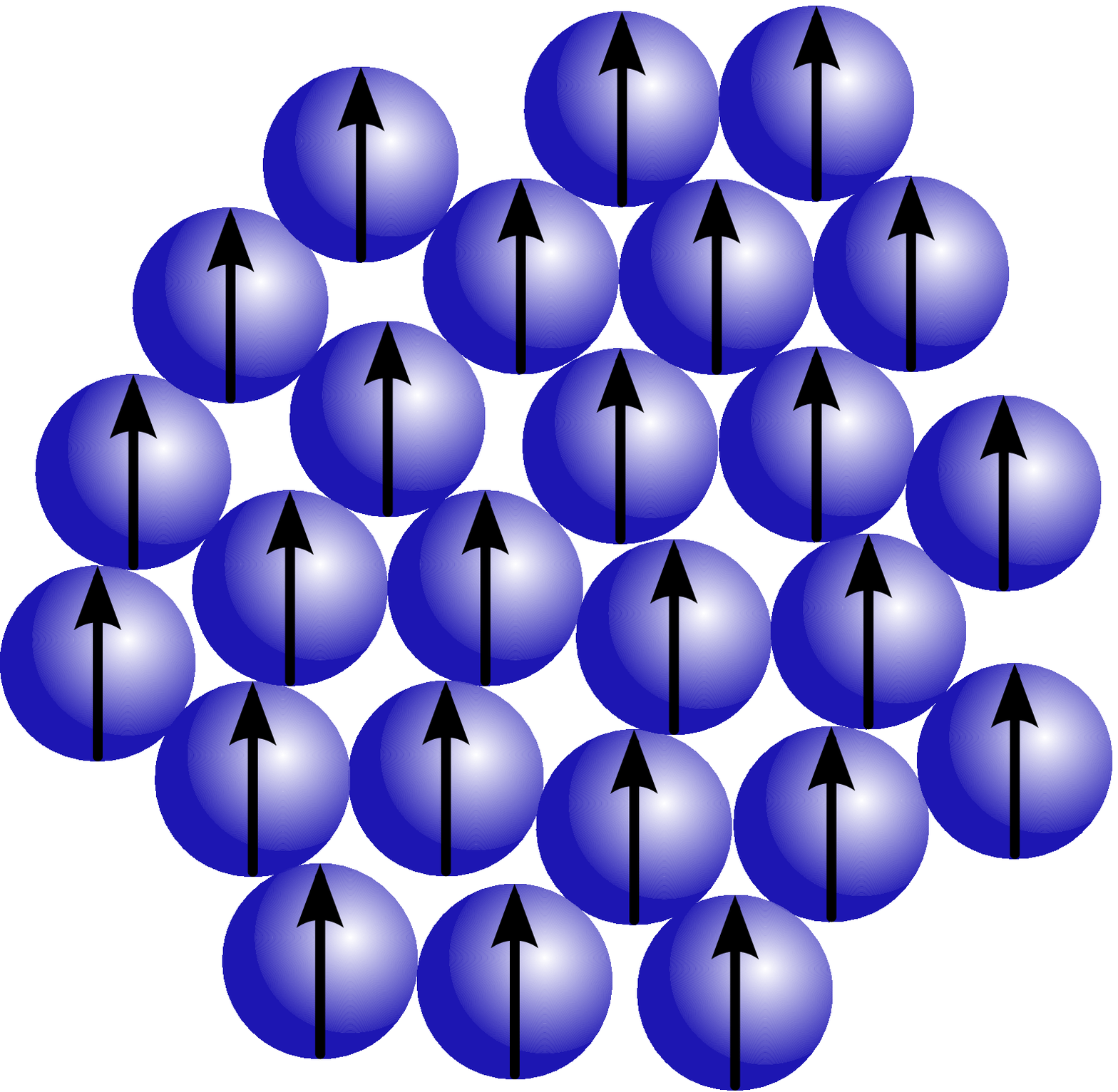}
    \label{fig.TotalForceScaling_all_align}%
  }%
  \hspace {0.05\columnwidth}
  \subfloat[][$\beta = \frac{d-1}{d} + \gamma$]{
    \centering
    \includegraphics[width = 0.4\columnwidth]{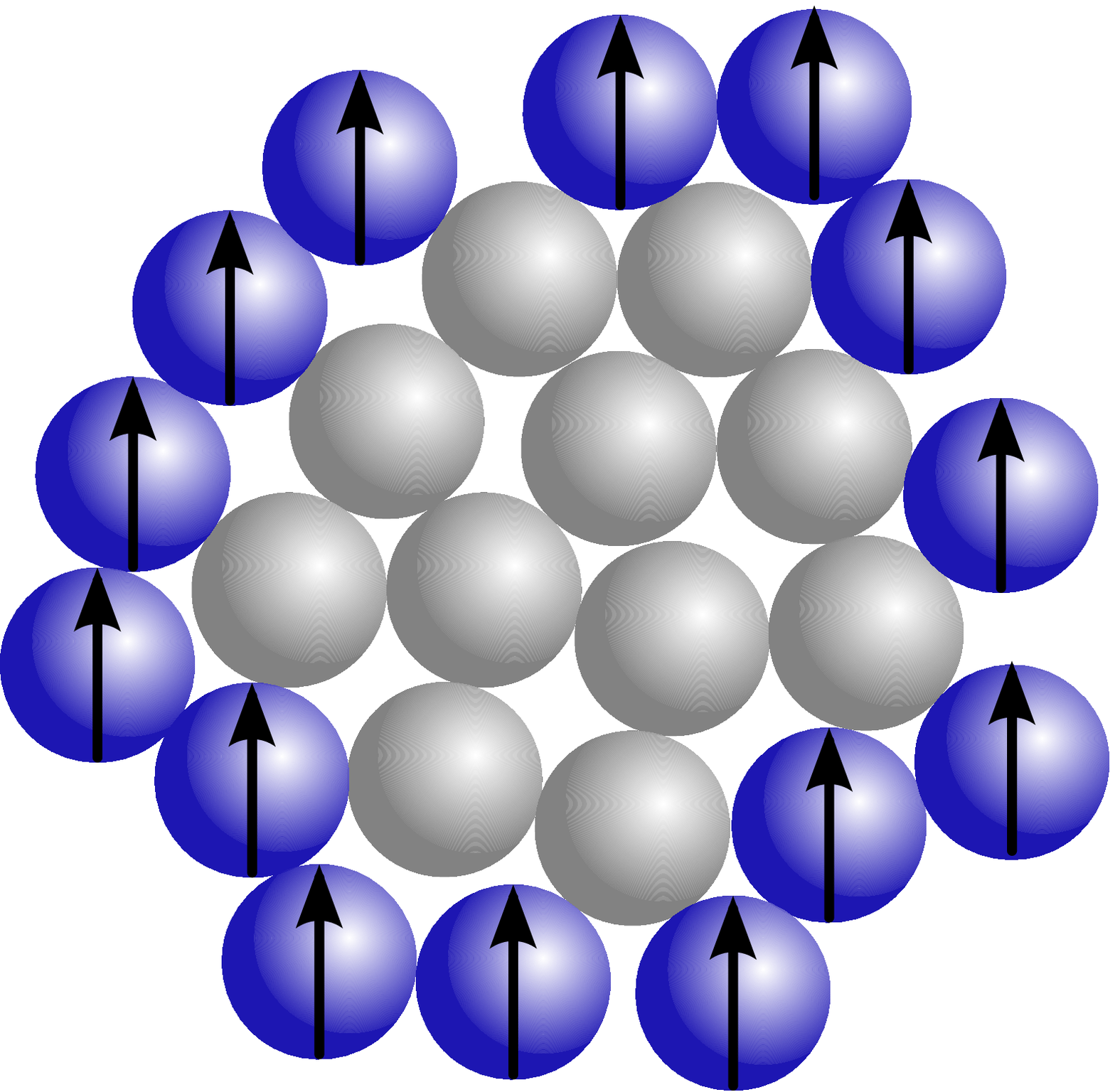}
    \label{fig.TotalForceScaling_surf_align}%
  }%
  \\
  \centering
  \subfloat[][$\beta = \frac{1}{2} + \gamma$]{
    \centering
    \includegraphics[width = 0.4\columnwidth]{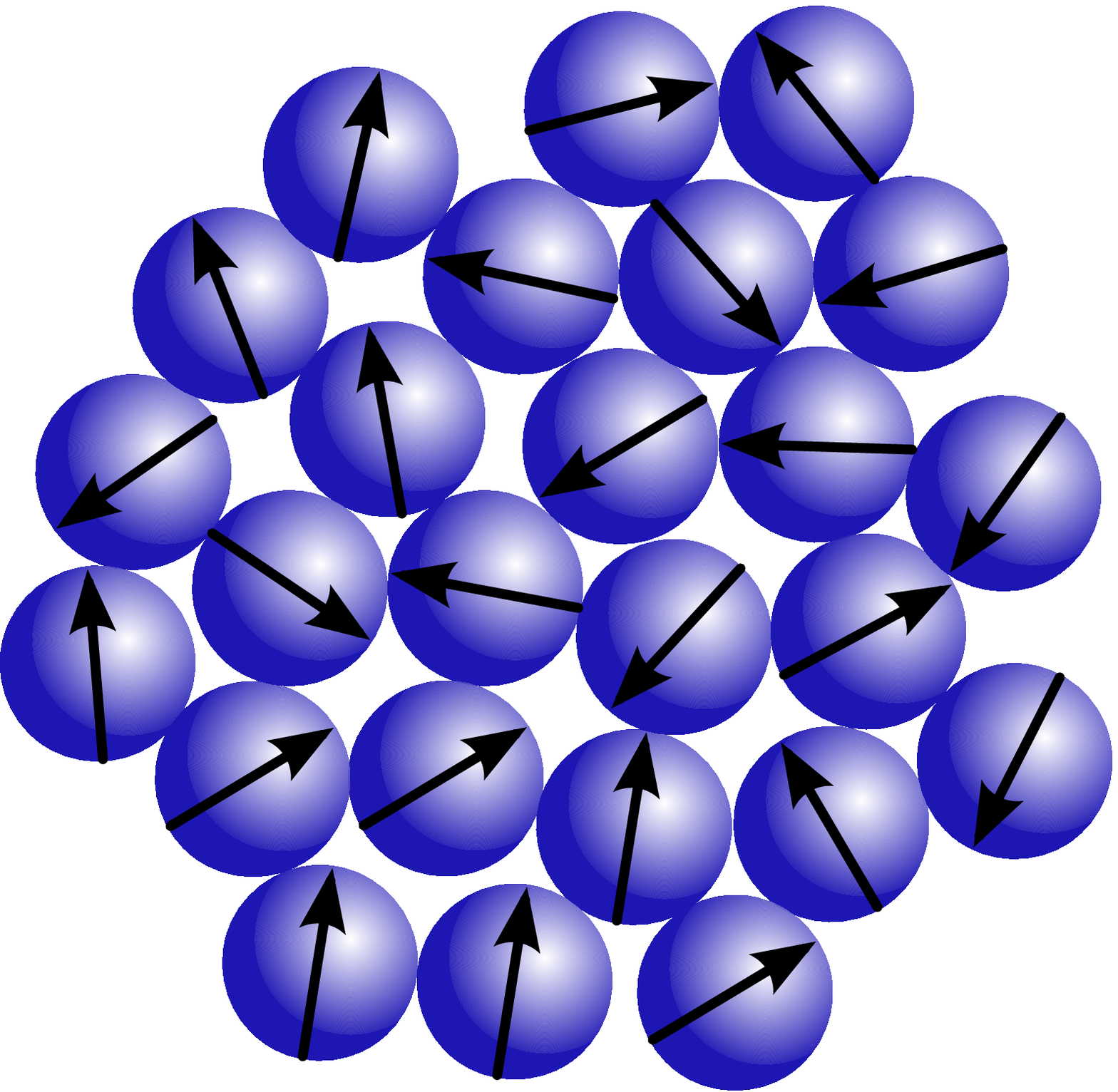}
    \label{fig.TotalForceScaling_all_random}%
  }%
  \hspace {0.05\columnwidth}
  \subfloat[][$\beta = \frac{d-1}{2d} + \gamma$]{
    \centering
    \includegraphics[width = 0.4\columnwidth]{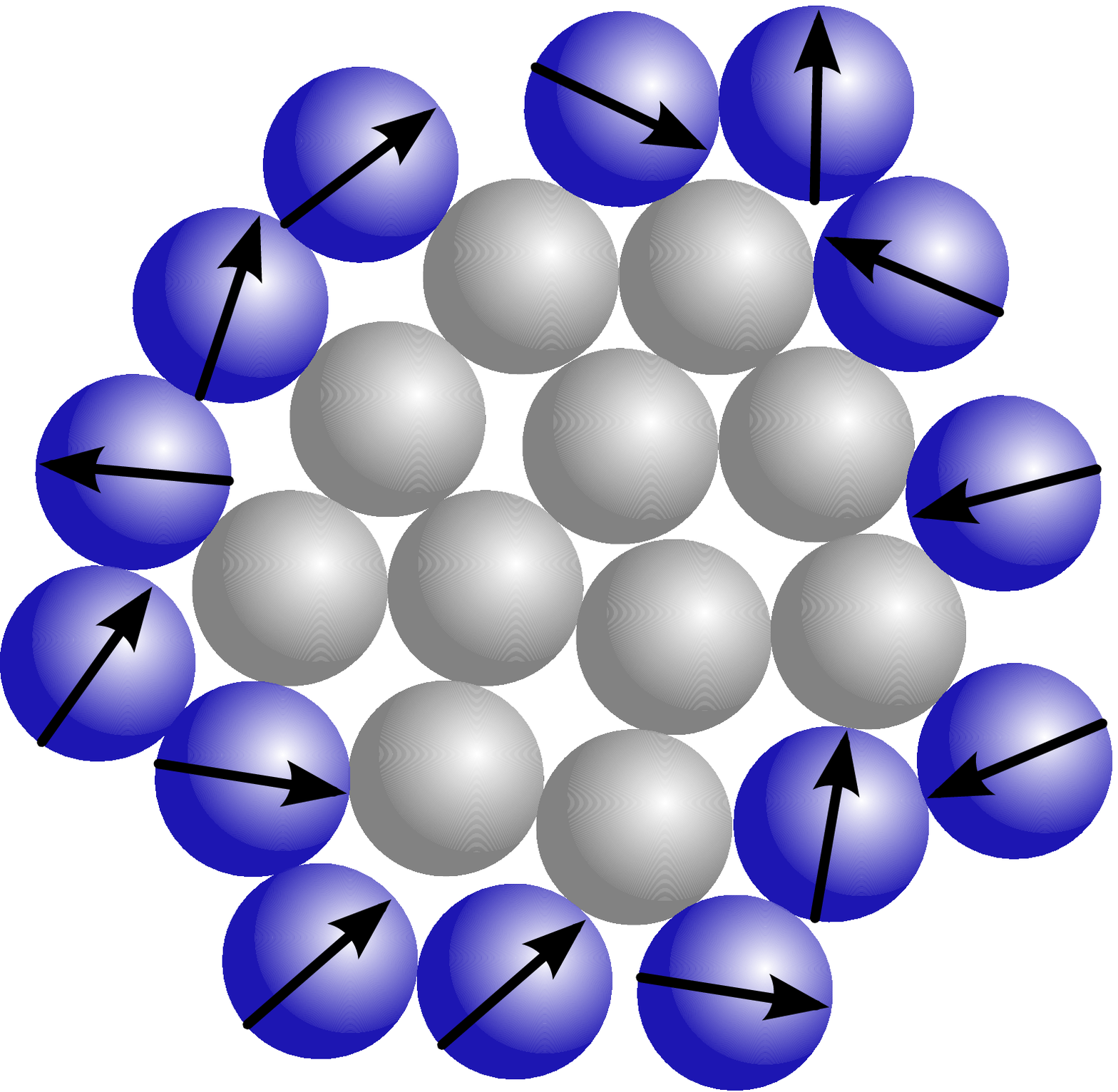}
    \label{fig.TotalForceScaling_surf_random}%
  }%
  \caption{Four cases for the scaling exponent $\beta$ of the total propulsion 
  force $F_N$ with cluster size $N$ in $d$ spatial dimensions. The arrows 
  denote the directions of single particle contribution forces.}
  \label{fig.TotalForceScaling}
\end{figure}

We now focus more on the individual forces $\mathbf{F}^{(i)}$ which 
constitute $F_N$. As discussed before, a fraction of the particles in a 
cluster can be rendered inactive, implying $\mathbf{F}^{(i)} = 0$ for all 
inactive particles. Apart from this we assume an additional overall reduction 
of the nonvanishing $\mathbf{F}^{(i)}$ with the cluster size. We describe this 
reduction by assuming a further scaling law 
$\mathbf{F}^{(i)} \propto N^\gamma$ with a general exponent $\gamma$. The 
exponent $\gamma$ vanishes for pushers and pullers 
\cite{Cisneros2007_ExpFluids,Bekham2008_ApplPhysLett} and for surface tension 
driven self--propelled droplets \cite{Thutupalli2011_NewJPhys} 
\footnote{
Individual surface tension driven swimmers have a propagation 
velocity which scales with the square of the radius 
\cite{Thutupalli2011_NewJPhys}. If these droplet particles merge to a larger 
droplet when clustering, case~\protect\subref{fig.TotalForceScaling_all_align} 
is realised and this scaling carries over to a cluster such that 
$v_N \sim R_N^2$. This leads to $\gamma = 0$. However, if the clustering 
cannot be associated with a merging of the droplet particles, they behave like 
pushers or pullers. The latter are known to realise 
case~\protect\subref{fig.TotalForceScaling_surf_align} with the scaling 
$v_N \sim R_N$ \cite{Cisneros2007_ExpFluids} and thus again possess 
$\gamma = 0$ by definition.}
. However, there are other situations where the effective individual forces 
$\mathbf{F}^{(i)}$ of contributing particles depend on the cluster size $N$ 
such that an overall reduction is relevant. Nontrivial values for $\gamma$ 
can be estimated by relating the scaling of the velocity $v^{(0)}$  of an 
individual particle with its radius  $R^{(0)}$ to the scaling of $v_N$ with 
$R_N$ via Eq.~\eqref{eq.ClusterVelocity}. Phoretic particles in $d = 3$ are 
propelled by a gradient generated on surface sites and their velocity is 
usually independent of the particle radius in three dimensions 
\cite{Golestanian2007_NewJPhys}. Ideally, the contributions of surface sites 
are aligned parallel and add up. Hence, $v_N \sim N^{(d-2)/d + \gamma}$ should 
not depend on $R_N \sim N^{1/d}$ in this situation which yields 
$\gamma = -(d-2)/d$. Likewise, the velocity of phoretic particles in a 
fuel--scarce environment is known to depend inversely on the particle radius 
\cite{Ebbens2012_PRE}, implying $v_N \sim N^{-1/d}$ for aligned surface 
contributions and thus $\gamma = -(d-1)/d$. 

Let us discuss the previously introduced four cases 
(see Fig.~\ref{fig.TotalForceScaling}) in more detail. For each of the four 
cases, one can simply compute the exponent $\beta$ for any prescribed $\gamma$ 
as follows: We define a further exponent $\lambda$ which measures the 
particles contributing to the cluster propulsion such that 
$F_N \sim N^\lambda F^{(i)} \sim N^\lambda N^\gamma$. Insertion into 
Eq.~\eqref{eq.ClusterVelocity} yields $\beta = \lambda + \gamma$. 
Contribution of all particles in case~\subref{fig.TotalForceScaling_all_align} 
means that $\lambda = 1$, while the 
case~\subref{fig.TotalForceScaling_surf_align} where only surface particles 
contribute corresponds to $\lambda = \frac{d-1}{d}$. Random alignment of the 
particles imposes a factor $1/2$ leading to $\lambda = 1/2$ in 
case~\subref{fig.TotalForceScaling_all_random} and $\lambda = \frac{d-1}{2d}$ 
in case~\subref{fig.TotalForceScaling_surf_random}. These exponents are  
included in Fig.~\ref{fig.TotalForceScaling}. 

We now introduce a simple \emph{sweeping argument} for active particles leading to 
scaling laws for the cluster size as a function of time $t$. Consider a 
typical cluster of size $N$ travelling through the system which has a uniform 
number density $\bar{\rho}$ of particles on average, no matter whether they 
are members of small clusters or noncoagulated, individual particles. 
Therefore, any inhomogeneities and local fluctuations in the particle and 
cluster distribution are neglected \footnote{These approximations can be 
abandoned in a more sophisticated Smoluchowski coagulation equation approach 
\cite{Dongen1988_JStatPhys}, which leads to the same scaling laws. Still, 
scaling itself is an assumption in the Smoluchowski approach which demands 
further numerical tests.}. If $V(t)$ denotes the volume in $d$-dimensional 
space which is covered by a cluster of size $R_N$ moving with velocity $v_N$ 
during a time $t$, we assume that all individual particles in this volume are 
irreversibly swept by the cluster. Differentially in time this implies
\begin{equation}
  \frac{dN}{dt} = {\bar \rho}  \frac{dV}{dt} .
  \label{eq.ClusterGrowthRate}
\end{equation}

Two limiting cases can be discriminated. In the so--called \emph{ballistic} 
regime, the persistence is so high that the cluster trajectory appears 
straight on the length scale the cluster possesses itself such that the rate 
of the swept volume is 
$\frac{dV}{dt} \propto v_N {R_N}^{d-1}\sim N^{(d-2)/d + \beta}$. This will occur 
in any case if the cluster becomes so large that rotational diffusion is 
suppressed \cite{Vicsek1995_PRL,Zheng2013_PRE}. In the complementary  case the 
cluster moves \emph{diffusively}. Then the effective diffusion constant of a 
random walk with step velocity $v_N$ scales as 
$D_N \sim {v_N}^2  \sim N^{2\beta - 2/d}$, such that the volume swept out is 
given by \cite{Veshchunov2010_JEngThermophys} 
$\frac{dV}{dt}  \sim {R_N}^{d-2} D_N \sim N^{(d-4)/d + 2\beta}$. Insertion 
into Eq.~\eqref{eq.ClusterGrowthRate} yields ordinary differential equations 
for $N(t)$ leading to our main result:
\begin{equation}
  N(t) = \begin{cases}
    \left[{N_0}^{2/d - \beta} + C(2/d - \beta)t  \right]^\frac{1}{2/d -\beta}	& \beta < 2/d	, \\
    N_0\exp(Ct)	& \beta = 2/d	, \\
    C(\beta - 2/d) \left(t_c - t \right)^\frac{-1}{\beta - 2/d}	& \beta > 2/d	,
  \end{cases}
  \label{eq.BallisticScalingLaws}
\end{equation}
for the \emph{ballistic} regime, and
\begin{equation}
  N(t) = \begin{cases}
    \left[{N_0}^{4/d - 2\beta} + C(4/d - 2\beta)t  \right]^\frac{1}{4/d - 2\beta}	& \beta < 2/d	, \\
    N_0\exp(Ct)	& \beta = 2/d	,
  \end{cases}
  \label{eq.DiffusiveScalingLaws}
\end{equation}
for the \emph{diffusive} regime, where $N_0=N(t=0)$ is the initial cluster 
size and $C$ is a positive amplitude prefactor. The last case of 
Eq.~\eqref{eq.BallisticScalingLaws} corresponds to an explosive growth 
scenario, where the cluster size diverges after a finite time 
$t_c = \frac{{N_0}^{2/d - \beta}}{C(\beta - 2/d)}$. Please note that 
$\beta > 2/d$ is never realised in the diffusive regime, as the cluster size 
would explode, which necessarily puts the system into the ballistic regime.
When measuring size in terms of the cluster radius the algebraic growth 
exponents of $R(t) \propto t^\alpha$ are given by 
$\alpha = \frac{1}{2 - \beta d}$ in the ballistic regime and  
$\alpha = \frac{1}{4 - 2\beta d}$ in the diffusive regime, which can be very 
large when $\beta d$ is close to but below $2$.
\section{Simulation}
\label{sec.simulation}
Using computer simulation, we investigate the cluster growth for various 
values of $\beta$ and different persistence lengths of cluster trajectories. 
The scaling of the total cluster force with an exponent $\beta$ from 
Eq.~\eqref{eq.ClusterVelocity} is an input in the simulation. Nevertheless, 
the final scaling of the cluster size with time as predicted by 
Eqs.~\eqref{eq.BallisticScalingLaws} and \eqref{eq.DiffusiveScalingLaws} is 
not an input but an output. Therefore, this final scaling behaviour is tested 
by our simulations. Moreover, the crossover to a possible ultimate scaling for 
\emph{finite} clusters can be addressed and computed in a simulation. 

In detail, the particles and compact clusters in these simulations are 
modelled as spherical droplets with radius $R_N = R^{{(0)}} N^{1/d}$, so that 
the total volume of all member particles is conserved. Initially, single 
particles start at random positions in a periodic simulation box with velocity 
$v^{{(0)}}$ and random direction. Particle collision events are predicted and 
on contact, particles merge at their centre of mass, forming larger clusters 
which again merge when colliding. The velocity of clusters is assigned to 
$v_N = v^{(0)} N^{\beta - 1/d} $. To model changes in the travelling direction 
of clusters in a general way, we use the following approach. After a 
reorientation time step $\Delta t$, a deviation from the current cluster 
direction is sampled for each member particle and the new cluster direction is
taken as the average of all the member particle deviations. Then the collision 
events for the new time step are predicted. When two clusters 
merge, we weight each cluster with its number of member particles in the 
direction of the merged cluster. Since the averaging process is a biased 
random walk, the persistence length of cluster trajectories increases with 
cluster size $N$. We sample the direction deviations of each particle from a 
von Mises--Fisher distribution \cite{Fisher1987_book} with concentration 
parameter $\kappa$, which is used as an input parameter. This distribution 
plays the role of the normal distribution on the $d$--dimensional unit sphere 
and $\kappa$ is similar to an inverse variance and determines the persistence 
of the trajectories such that we refer to $\kappa$ as the persistence 
parameter in the following. Both, the ballistic and diffusive regime, can be 
gained as extreme limits $\kappa \to\infty$ or $\kappa =0$.

\begin{figure}[htb]
  \setlength{\fboxsep}{0pt}%
  \centering
  \subfloat[][$t = 0$]{
    \centering
    \fbox{%
      \includegraphics[width = 0.48\columnwidth]{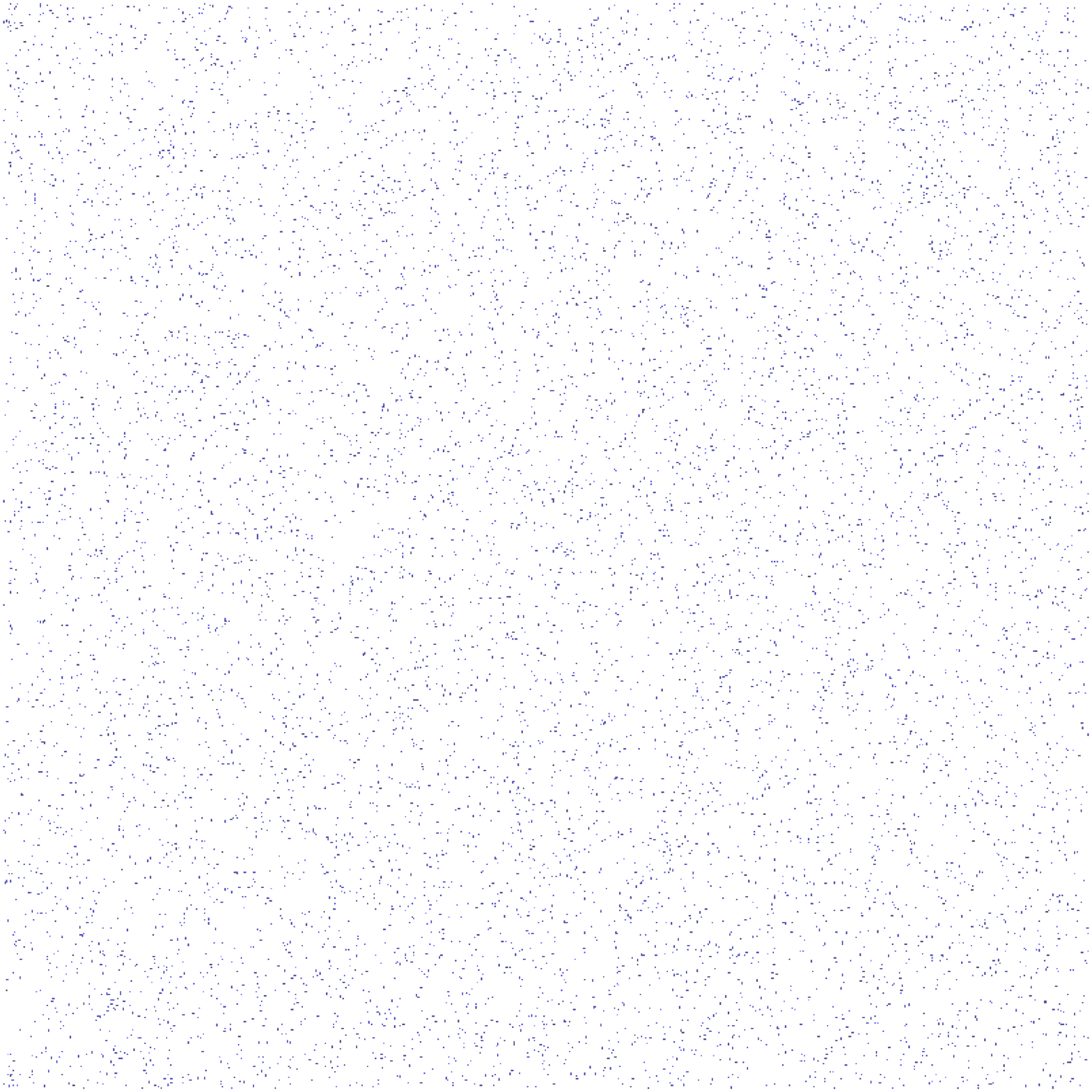}%
    }%
    \label{fig.snapshot_0}%
  }%
  \subfloat[][$t = 71.6\tau$]{
    \centering
    \fbox{%
      \includegraphics[width = 0.48\columnwidth]{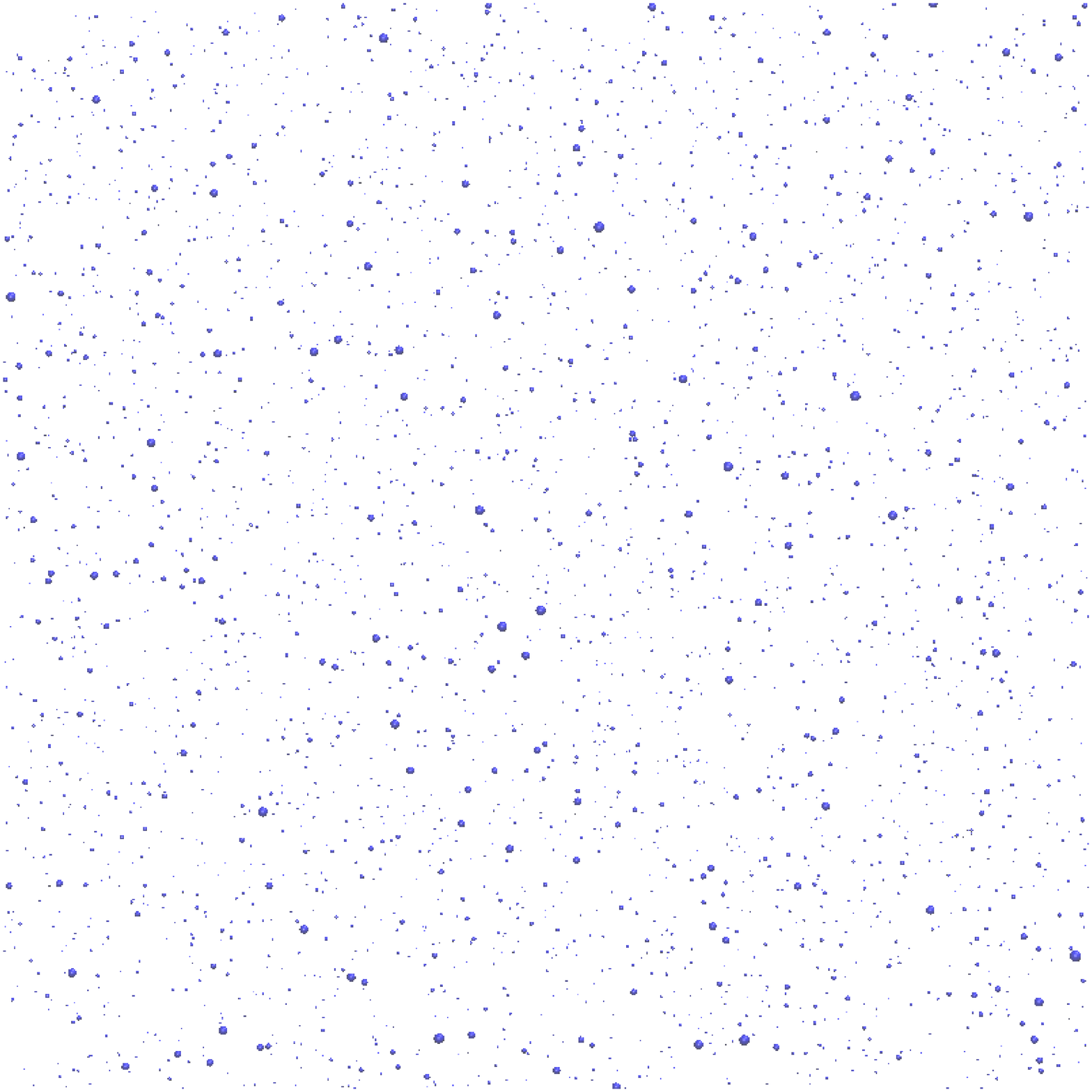}%
    }%
    \label{fig.snapshot_716}%
  }%
  \\
  \centering
  \subfloat[][$t = 143.2\tau$]{
    \centering
    \fbox{%
      \includegraphics[width = 0.48\columnwidth]{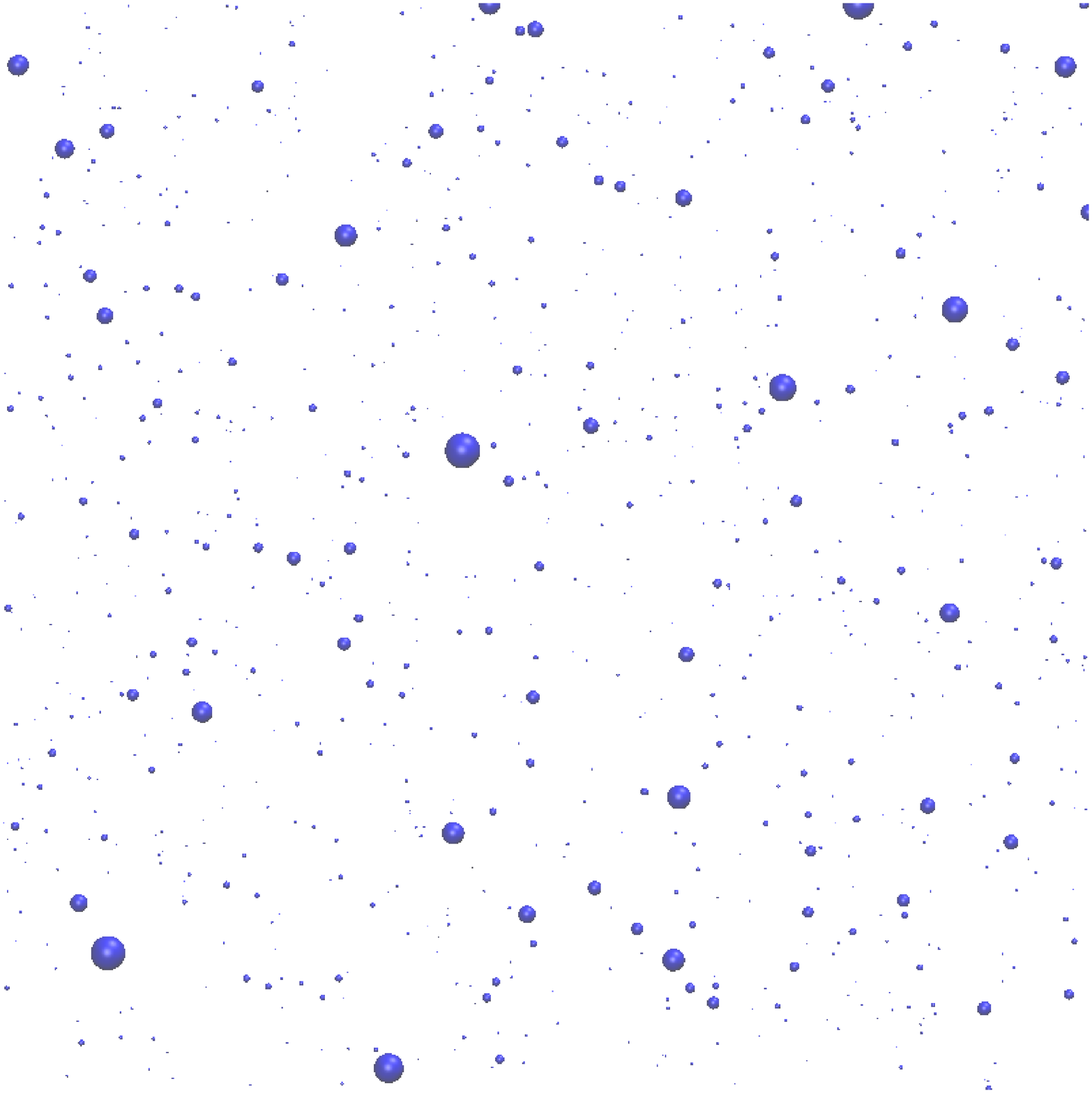}%
    }%
    \label{fig.snapshot_1432}%
  }%
  \subfloat[][$t = 214.8\tau$]{
    \centering
    \fbox{%
      \includegraphics[width = 0.48\columnwidth]{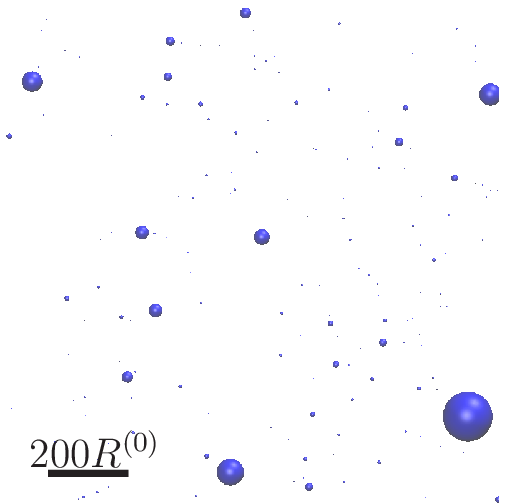}%
    }%
    \label{fig.snapshot_2148}%
  }%
  \caption{Typical simulation snapshots at various times, here in a 
  two--dimensional   simulation of $\mathcal{N} = 10^4$ total particles at 
  packing fraction $\eta = 0.02$ with $\beta = 1$ and persistence parameter 
  $\kappa = 100$. Snapshot \protect\subref{fig.snapshot_2148} shows the 
  situation just before the simulation is terminated.}%
  \label{fig.snapshots}
\end{figure}

The single particle radius $R^{(0)}$ defines the length scale in the system, 
while the time $\tau = R^{(0)}/v^{(0)}$ a single particle requires to travel 
its own radius is used as time scale. We chose $\Delta t = 0.1 \tau$, which is 
sufficiently small when using a collision event prediction scheme. The packing 
fraction is taken as $\eta = 0.02$ with $\mathcal{N} = 10^6$ initial 
particles. We vary the scaling exponent $\beta$ of the total propulsion force 
as well as the persistence parameter $\kappa$ in $d = 2,3$ dimensions. Since 
at late times the system is depleted of particles and the anticipated scaling 
laws clearly cannot be observed any more, we terminate the simulation as soon 
as a cluster reaches a size of $R_N > 0,05 L$, where $L$ is the box length. 
Typical snapshots from a simulation are shown in Fig.~\ref{fig.snapshots}.

\begin{figure*}[tbp]
  \subfloat[][$\kappa = 0$, $d = 2$]{
    \centering
    \includegraphics[scale = 0.9]{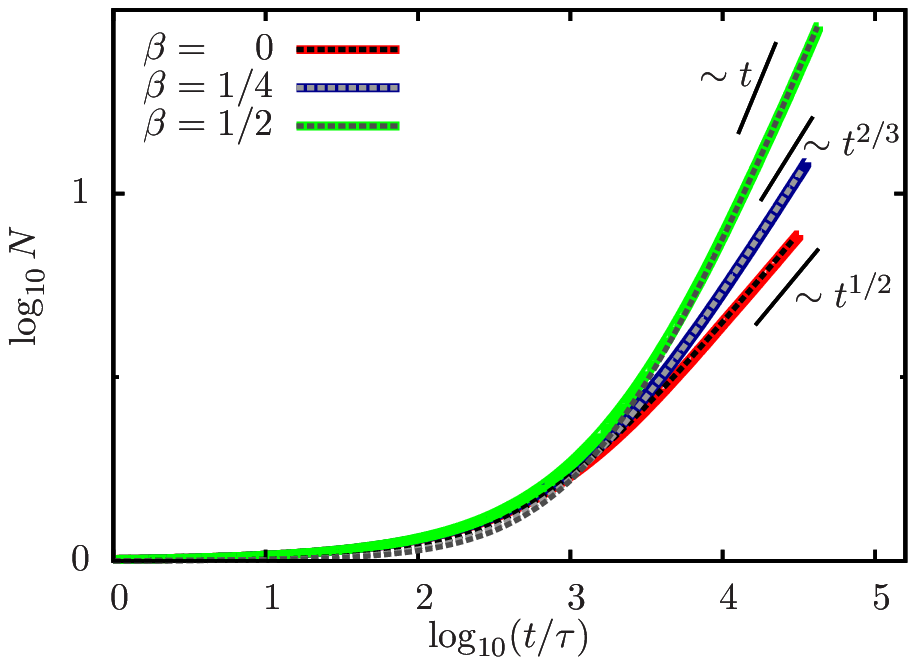}%
    \label{fig.DiffusiveAlgebraic_2d}%
  }%
  \subfloat[][$\kappa = 0$, $d = 3$]{
    \centering
    \includegraphics[scale = 0.9]{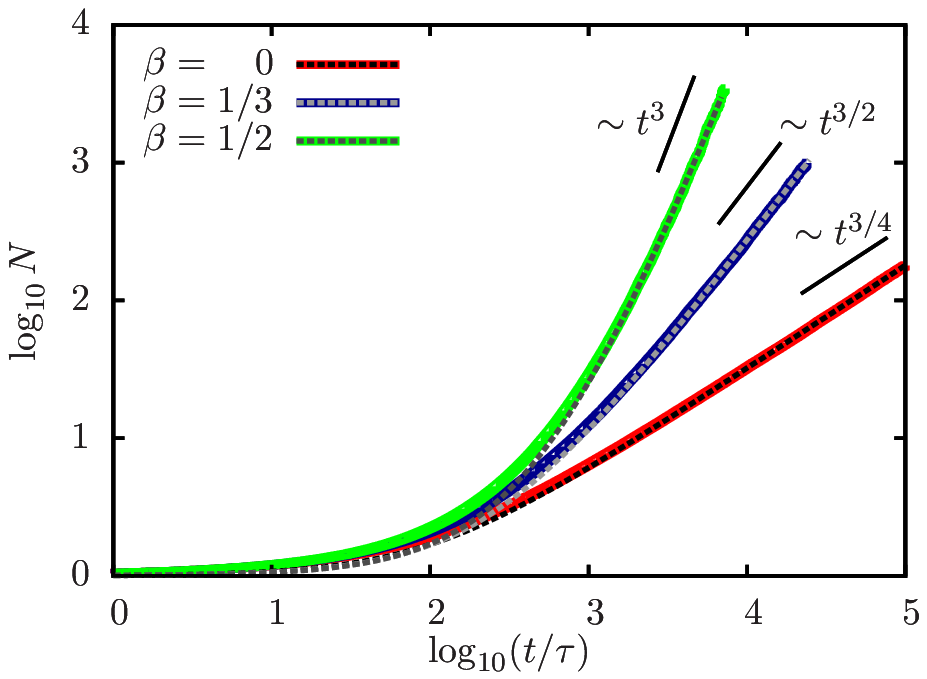}%
    \label{fig.DiffusiveAlgebraic_3d}%
  }%
  \\
  \subfloat[][$\kappa = 100$, $d = 2$]{
    \centering
    \includegraphics[scale = 0.9]{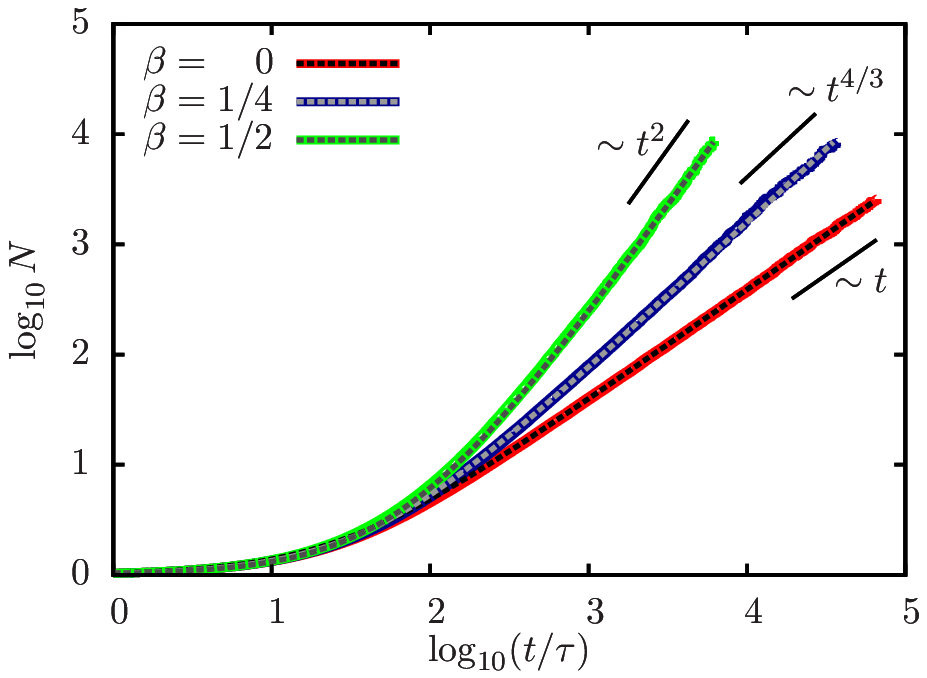}%
    \label{fig.BallisticAlgebraic_2d}%
  }%
  \subfloat[][$\kappa = 100$, $d = 3$]{
    \centering
    \includegraphics[scale = 0.9]{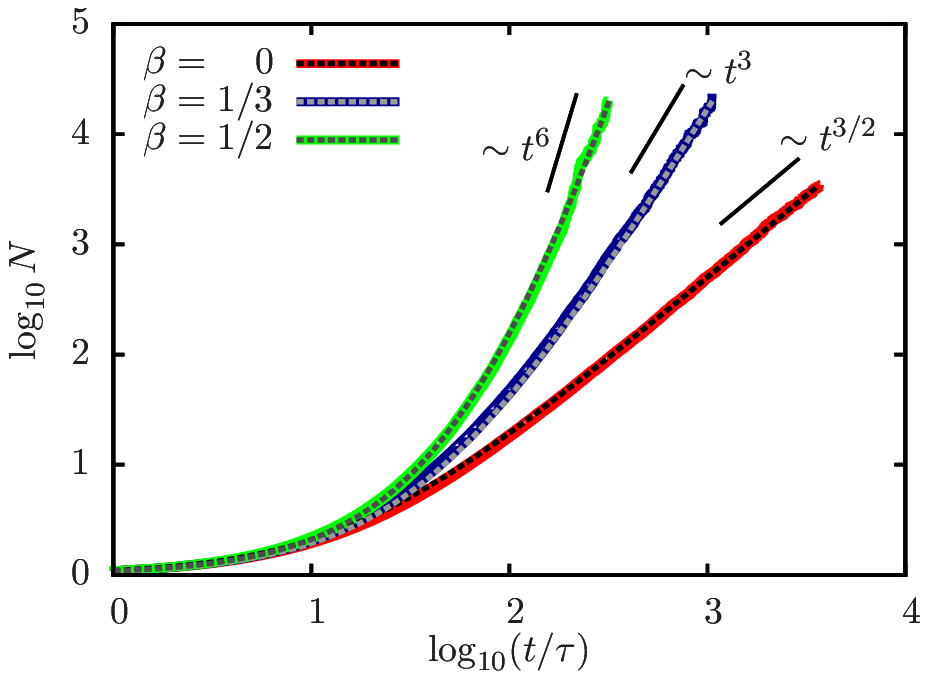}%
    \label{fig.BallisticAlgebraic_3d}%
  }%
  \\
  \subfloat[][$\beta = 2/3$, $d = 3$]{
    \centering
    \includegraphics[scale = 0.9]{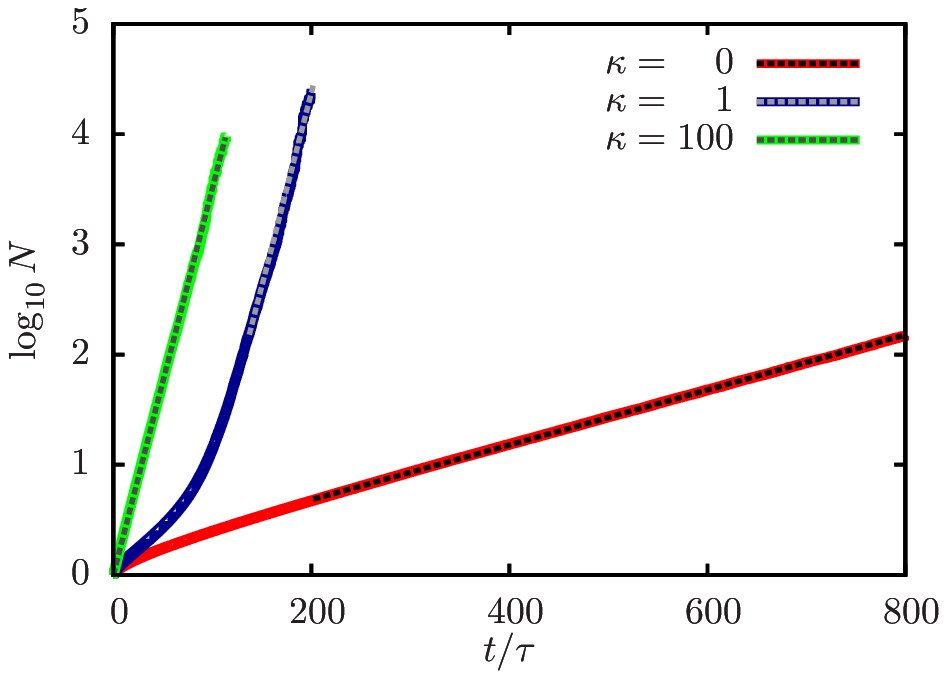}%
    \label{fig.Exponential_3d}%
  }%
  \subfloat[][$\beta = 1$, $\kappa = 100$, $d = 3$]{
    \centering
    \includegraphics[scale = 0.9]{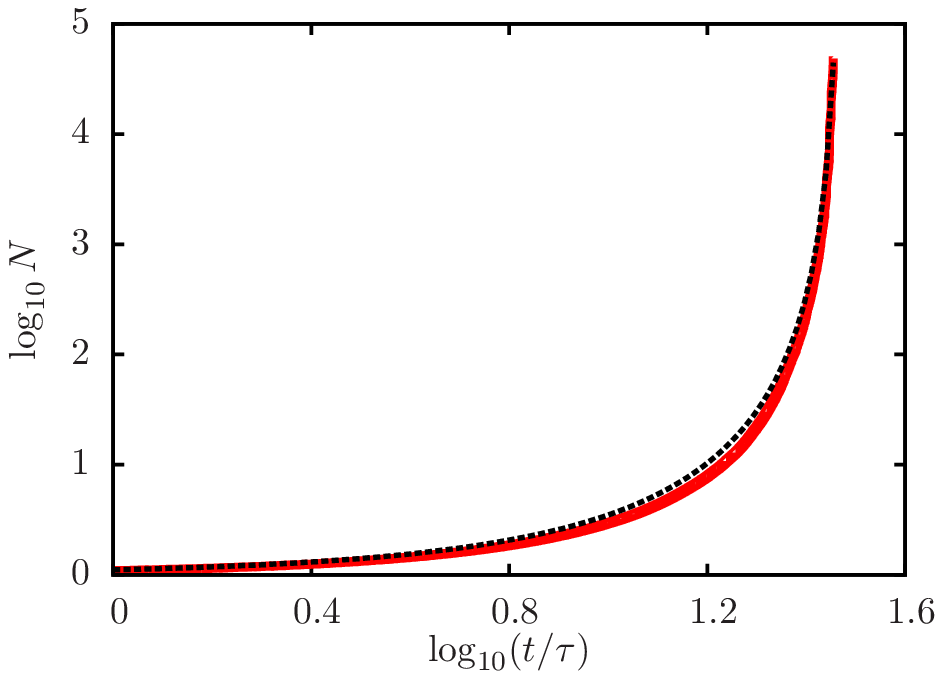}%
    \label{fig.Explosive}%
  }%
  \caption{Cluster size evolution obtained from simulations in $d = 2,3$ 
  dimensions with various   values of the persistence parameter $\kappa$ and 
  the total propulsion force scaling exponent   $\beta$. Algebraic growth in 
  the diffusive regime  \protect\subref{fig.DiffusiveAlgebraic_2d}, 
  \protect\subref{fig.DiffusiveAlgebraic_3d} as well as in the ballistic 
  regime \protect\subref{fig.BallisticAlgebraic_2d}, 
  \protect\subref{fig.BallisticAlgebraic_3d} occurs with the predicted 
  exponents as indicated in the plots. Exponential growth 
  \protect\subref{fig.Exponential_3d} in the ballistic regime occurs faster 
  than in the diffusive regime as indicated by the much higher slope. 
  For high persistence and high force scaling, explosive cluster growth occurs  
  \protect\subref{fig.Explosive}. The dashed lines are fits using 
  Eqs.~\eqref{eq.BallisticScalingLaws} and \eqref{eq.DiffusiveScalingLaws} 
  respectively.}%
  \label{fig.data}
\end{figure*}

Figures~\ref{fig.DiffusiveAlgebraic_2d} and \ref{fig.DiffusiveAlgebraic_3d} 
show the evolution of the mean cluster size $N(t)$ in two-- and 
three--dimensional simulations in the diffusive regime ($\kappa =0$) for 
various values of $\beta < 2/d$. Fits of $N(t)$ with 
Eq.~\eqref{eq.DiffusiveScalingLaws} possess the predicted algebraic scaling. 
Data in the ballistic regime are shown in 
Figs.~\ref{fig.BallisticAlgebraic_2d} and \ref{fig.BallisticAlgebraic_3d}. The 
predicted scaling exponent for the algebraic growth is verified for all values 
of $\beta$.

For the case of $\beta = 2/d$, the predicted exponential growth in both 
regimes is reproduced by the simulations and the prefactor $C$ in the 
ballistic regime is significantly larger than in the diffusive regime, see the 
slope of the semilogarithmic plots in Fig.~\ref{fig.Exponential_3d}. The slope 
of the plot for $\kappa = 1$ steadily increases until it reaches the level for 
$\kappa = 100$ as clusters in this system need to grow first to enter the 
ballistic regime. Finally, for $\beta > 2/3$ in three dimensions explosive 
cluster growth is documented in Fig.~\ref{fig.Explosive} which confirms the 
predicted scaling $\sim (t_c - t)^{-3}$.

\section{Fractal aggregates}
\label{sec.fractals}
Our results can be further extended to clusters which lack the reorganisation 
mechanism leading to a compact shape. When particles and clusters simply stick 
to each other on the first point of contact, the resulting shapes possess a 
ramified and fractal structure \cite{Meakin1984_PRA,Kolb1983_PRL}.

The size of such aggregates can be conveniently described by the radius 
of gyration $R_N^{(g)}$ which replaces $R_N$ and is approximately proportional 
to the hydrodynamic radius $R_N^{(h)}$ 
\cite{Saarloos1987_PhysicaA,Lattuada2003_JColloidInterfaceSci}. A structure 
with fractal dimension $d_F$ ($1 \leq d_F \leq d$) then implies the scaling
$R_N^{(g)} \sim N^{1/d_F}$. Therefore, the analogue to 
Eq.~\eqref{eq.ClusterVelocity} for fractal clusters is
\begin{equation}
  v_N \sim F_N/R_N^{(h)} \sim F_N/R_N^{(g)} \sim N^{\beta - 1/d_F}.
  \label{eq.ClusterVelocity_fractal}
\end{equation}

\begin{figure}[htb]
  \setlength{\fboxsep}{0pt}%
  \centering
  \subfloat[][$t = 0$]{
    \centering
    \fbox{%
      \includegraphics[width = 0.48\columnwidth]{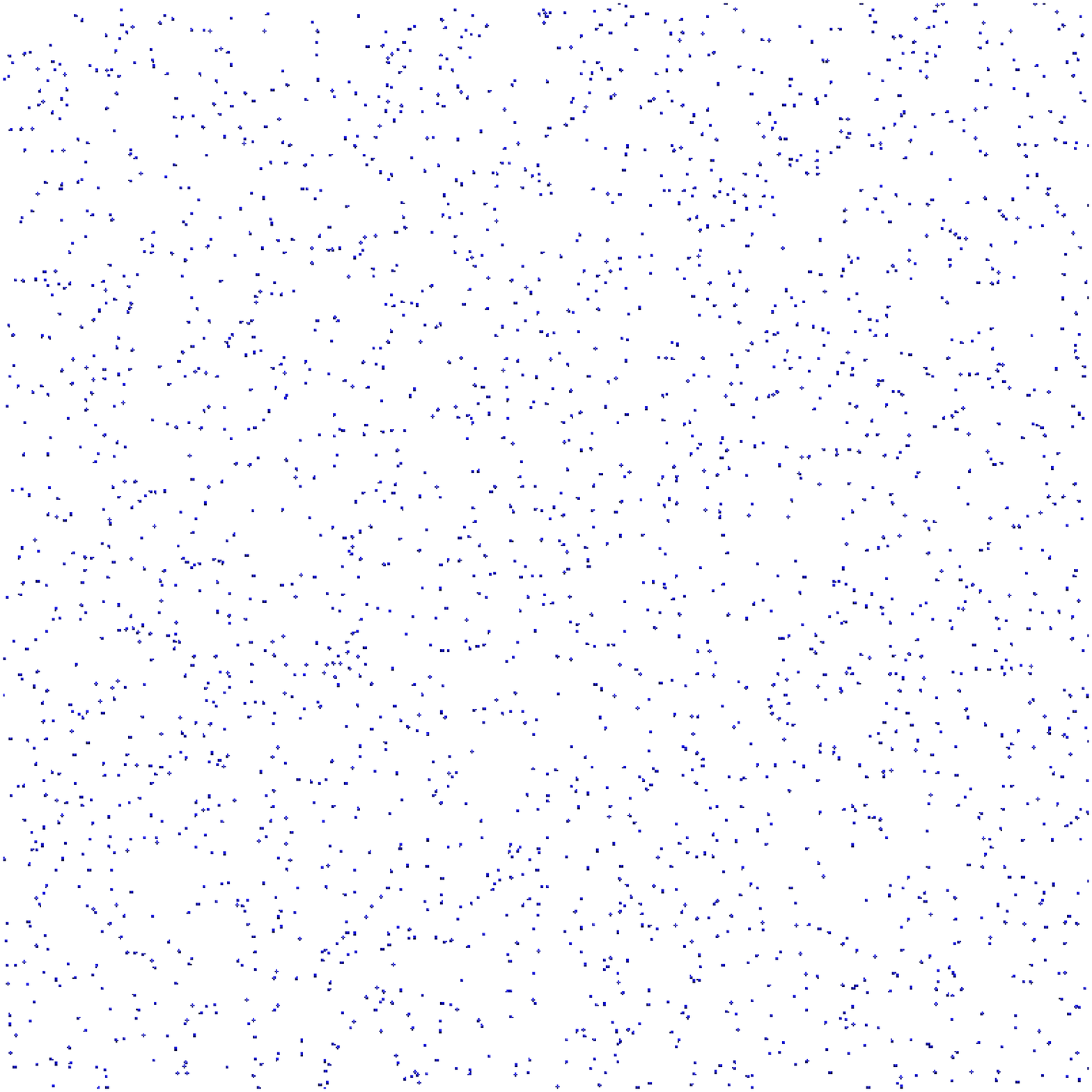}%
    }%
    \label{fig.snapshot_fractal_0}%
  }%
  \subfloat[][$t = 40\tau$]{
    \centering
    \fbox{%
      \includegraphics[width = 0.48\columnwidth]{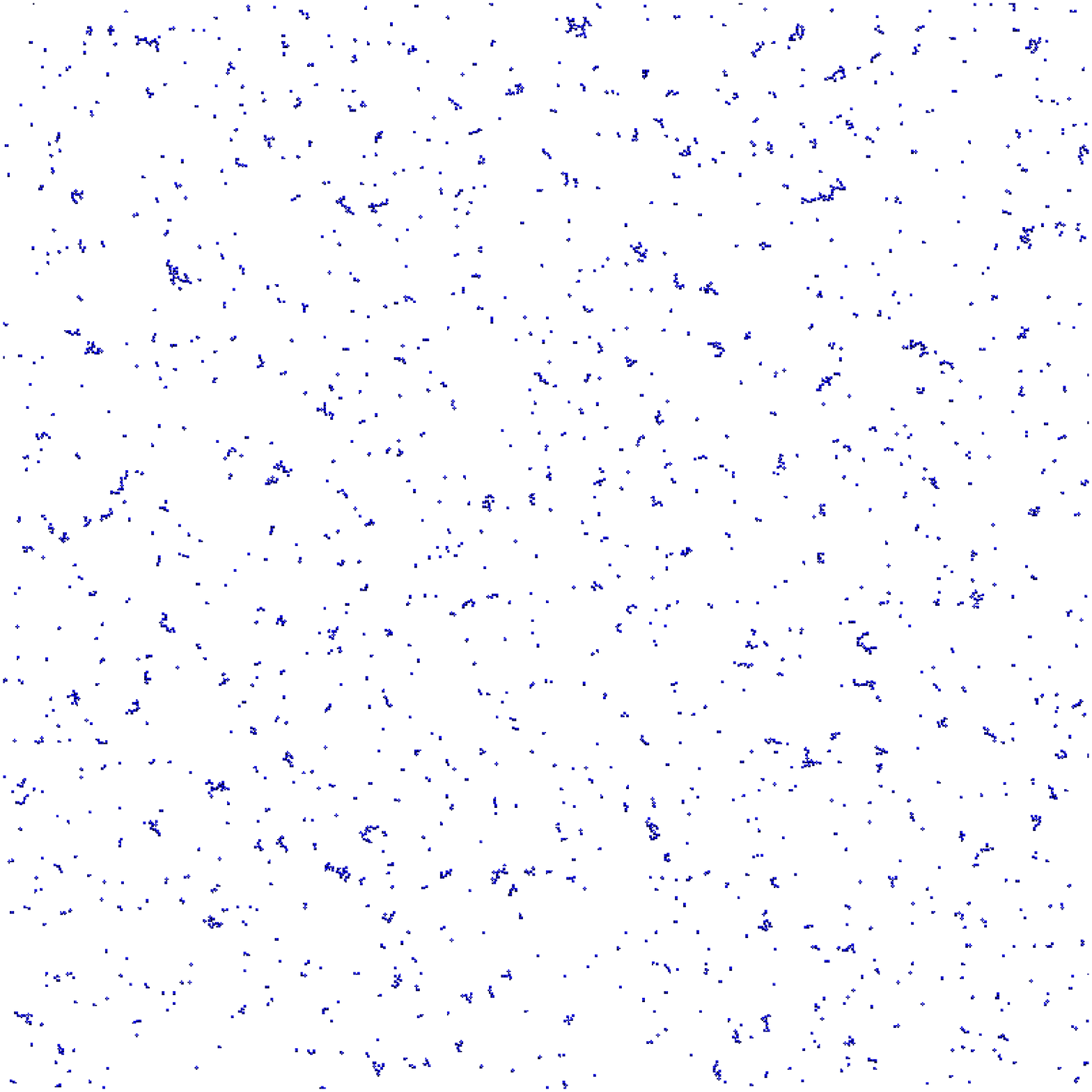}%
    }%
    \label{fig.snapshot_fractal_40}%
  }%
  \\
  \centering
  \subfloat[][$t = 80\tau$]{
    \centering
    \fbox{%
      \includegraphics[width = 0.48\columnwidth]{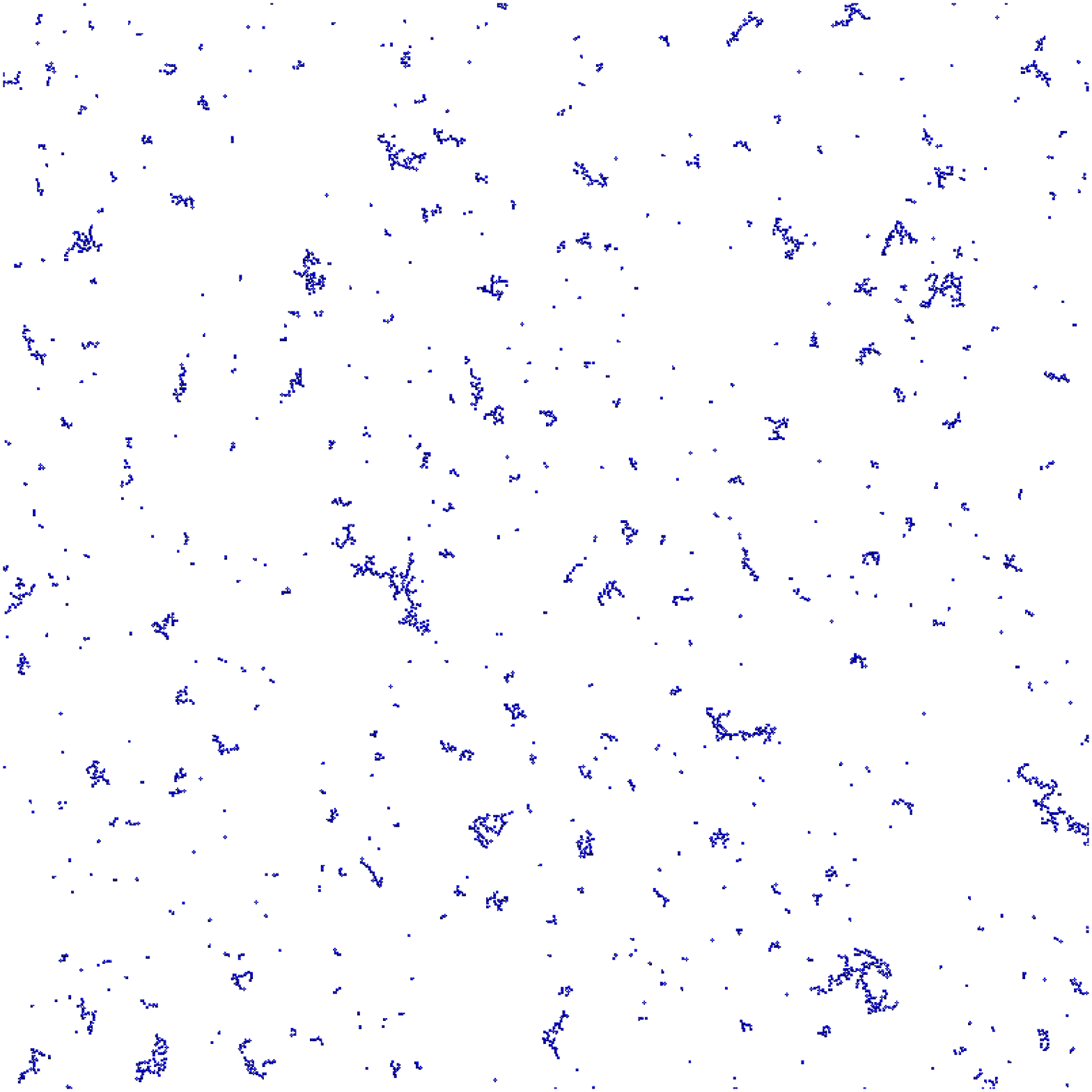}%
    }%
    \label{fig.snapshot_fractal_80}%
  }%
  \subfloat[][$t = 121\tau$]{
    \centering
    \fbox{%
      \includegraphics[width = 0.48\columnwidth]{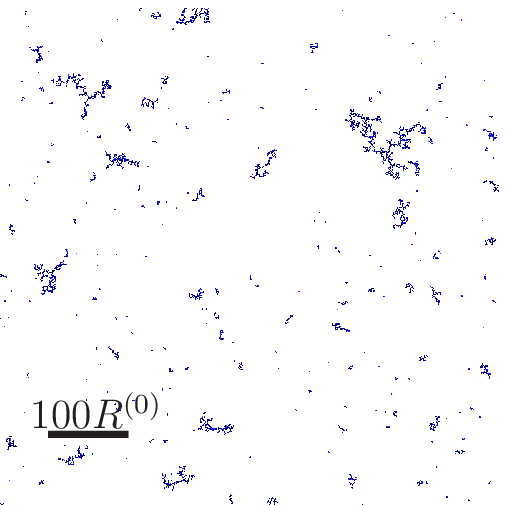}%
    }%
    \label{fig.snapshot_fractal_121}%
  }%
  \caption{Typical snapshots of a two--dimensional simulation of fractal clustering using the same parameters as in Fig.~\ref{fig.snapshots}, 
  except for the persistence parameter $\kappa$ which is now chosen to be $\kappa = 10$.}%
  \label{fig.snapshots_fractal}
\end{figure}

\begin{figure*}[tbp]
  \centering
  \subfloat[][$\kappa = 100$, $d = 2$]{
    \centering
    \includegraphics[scale = 0.9]{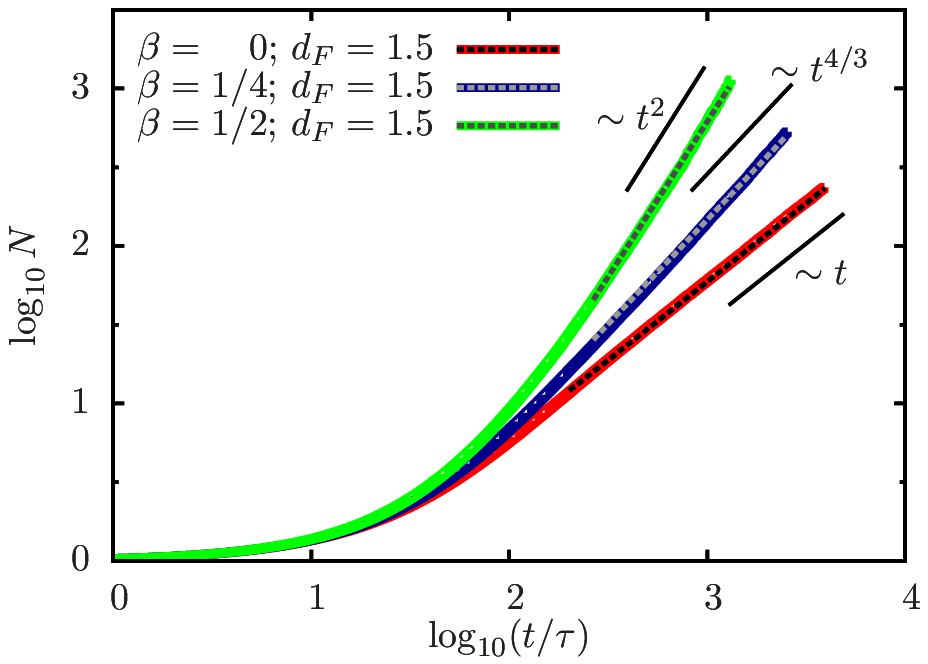}%
    \label{fig.BallisticAlgebraic_2d_fractal}%
  }%
  \subfloat[][$\beta = 1$, $d = 2$]{
    \centering
    \includegraphics[scale = 0.9]{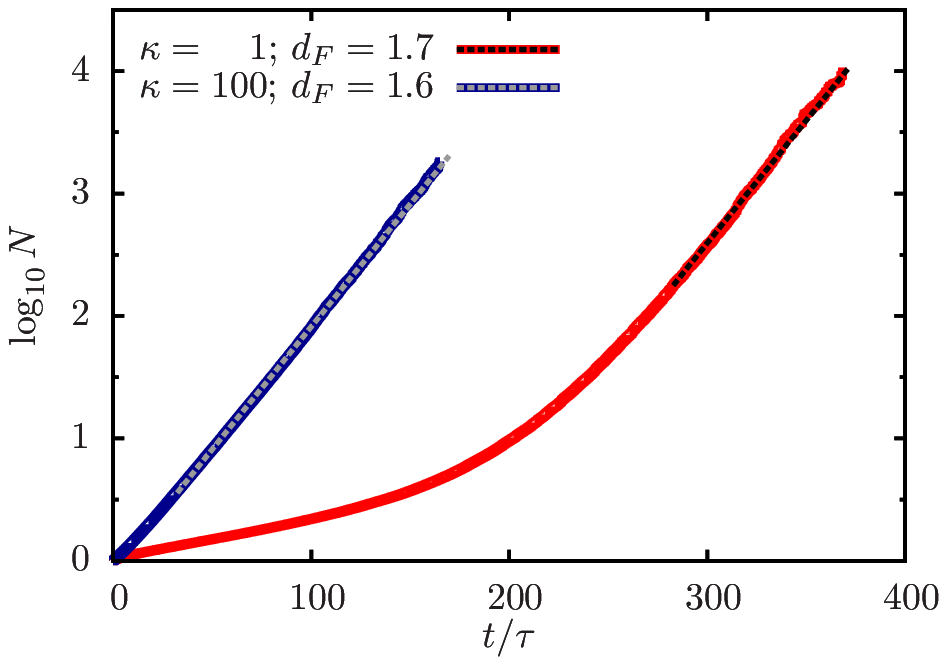}%
    \label{fig.Exponential_2d_fractal}%
  }%
  \\
  \subfloat[][$\kappa = 0$, $d = 2$]{
    \centering
    \includegraphics[scale = 0.9]{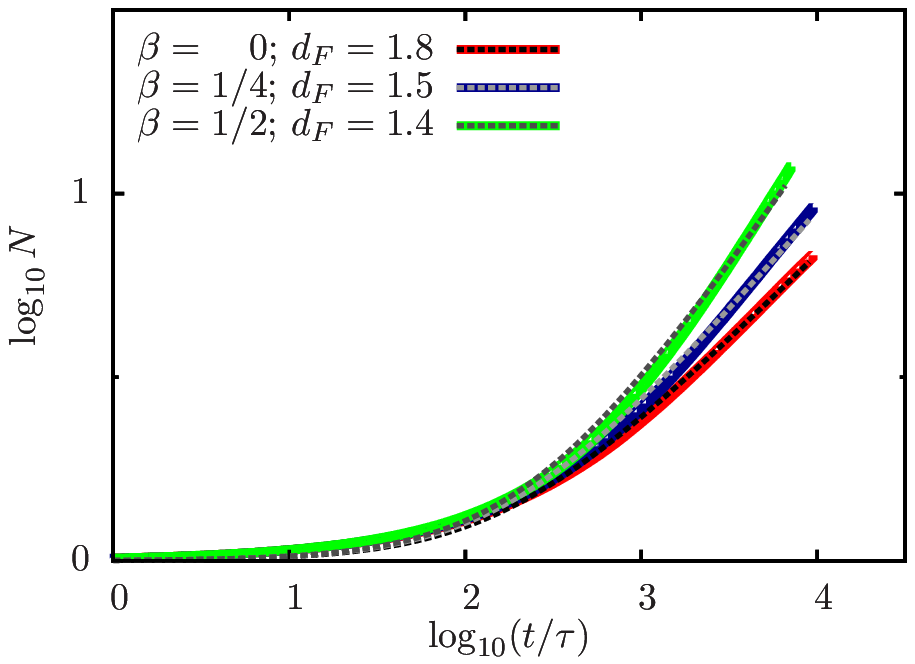}%
    \label{fig.DiffusiveAlgebraic_2d_fractal}%
  }%
  \subfloat[][$\kappa = 100$, $d = 3$]{
    \centering
    \includegraphics[scale = 0.9]{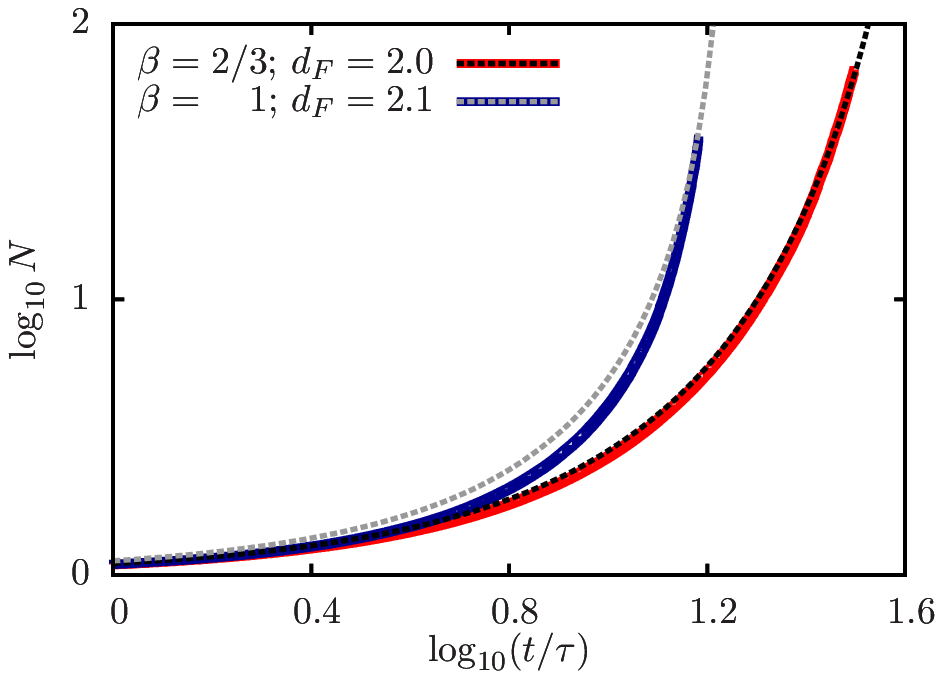}%
    \label{fig.Explosive_fractal}%
  }%
  \caption{Fractal cluster size evolution obtained from simulations in 
  $d = 2,3$ dimensions with various values of the persistence parameter 
  $\kappa$ and the total propulsion force scaling exponent $\beta$. For 
  $d = 2$ in the ballistic regime 
  \protect\subref{fig.BallisticAlgebraic_2d_fractal}, the same scaling laws 
  hold as in the compact case. Accordingly, the threshold for exponential 
  growth in the ballistic regime \protect\subref{fig.Exponential_2d_fractal} 
  remains at $\beta = 1$. In the diffusive regime for $d = 2$ 
  \protect\subref{fig.DiffusiveAlgebraic_2d_fractal} the fractal dimension 
  $d_F$ enters the scaling laws and leads to slower growth as compared to the 
  compact case. Thresholds in $\beta$ for explosive growth are lower for 
  $d_F < d$ so that such growth behaviour occurs already for $\beta = 2/3$ 
  when $d = 3$ \protect\subref{fig.Explosive_fractal}. The dashed lines are 
  fits using Eqs.~\eqref{eq.BallisticScalingLaws_fractal} and 
  \eqref{eq.DiffusiveScalingLaws_fractal} respectively.}%
  \label{fig.data_fractal}
\end{figure*}

We have performed additional simulations implementing irreversible sticking of 
particles at the point of contact. Equation~\eqref{eq.ClusterVelocity_fractal} 
is taken into account by assigning the cluster velocity to 
$v_N = v^{(0)} N^\beta / R_N^{(g)}$. Therefore, the radius of gyration of each 
cluster has to be tracked throughout the simulation. Apart from this, the 
simulation follows the same procedure as in \ref{sec.simulation}. 
Figure~\ref{fig.snapshots_fractal} shows typical snapshots confirming the 
ramified structure of the aggregates. We have determined the fractal dimension 
$d_F$ from the simulation data for the cluster structure. Results for $d_F$ 
are presented in the legends of Fig.~\ref{fig.data_fractal}.

In addition to the drag, the radius of gyration also determines the collision 
cross--section of the cluster. Applying the same simple sweeping argument used 
in \ref{sec.model}, we obtain 
$\frac{dV}{dt} \propto v_N {R_N^{(g)}}^{d-1} \sim N^{(d - 2)/d_F + \beta}$ 
for the rate of the swept volume in the ballistic regime and 
$\frac{dV}{dt} \sim {R_N^{(g)}}^{d-2} D_N \sim N^{(d-4)/d_F + 2\beta}$
in the diffusive regime. Obviously, the sweeping volumes of compact clusters 
are recovered when setting $d_F = d$. Insertion into 
Eq.~\eqref{eq.ClusterGrowthRate} then yields the scaling relations 
\begin{equation}
  N(t) = \begin{cases}
    \left[{N_0}^{\xi_b - \beta} + C(\xi_b - \beta)t  \right]^\frac{1}{\xi_b - \beta}	& \beta < \xi_b	, \\
    N_0\exp(Ct)	& \beta = \xi_b	, \\
    C(\beta - \xi_b) \left(t_c - t \right)^\frac{-1}{\xi_b - \beta}	& \beta > \xi_b	,
  \end{cases}
  \label{eq.BallisticScalingLaws_fractal}
\end{equation}
for the \emph{ballistic} regime where the abbreviation 
$\xi_b = (2 - d)/d_F + 1$ is used. The critical time for explosive growth is 
$t_c = \frac{{N_0}^{\xi_b - \beta}}{C(\beta - \xi_b)}$ here. Note, that for 
$d = 2$ these results are indistinguishable from 
Eq.~\eqref{eq.BallisticScalingLaws} as $\xi_b$ does not depend on $d_F$ for 
$d = 2$. 

For the \emph{diffusive} regime we obtain the scaling law
\begin{equation}
  N(t) = \begin{cases}
    \left[{N_0}^{\xi_d - 2\beta} + C(\xi_d - 2\beta)t  \right]^\frac{1}{\xi_d - 2\beta}	& \beta < \xi_d / 2	, \\
    N_0\exp(Ct)	& \beta = \xi_d / 2	, \\
  \end{cases}
  \label{eq.DiffusiveScalingLaws_fractal}
\end{equation}
with $\xi_d = (4 - d)/d_F + 1$. Contrary to the behaviour 
in the ballistic regime, the threshold value for $\beta$ corresponding to 
exponential growth is raised as compared to compact clusters. Similarly, the 
algebraic growth exponents are lower for the same value of $\beta$. 

Computer simulation results verifying the growth behaviour for $d = 2$ in the 
ballistic regime are shown in Fig.~\ref{fig.BallisticAlgebraic_2d_fractal} for 
the case of algebraic growth as well as in 
Fig.~\ref{fig.Exponential_2d_fractal} for exponential growth. In fact, for 
$d = 2$, the algebraic growth exponents and the threshold for exponential 
growth are the same as in the compact case. Conversely, for $d = 3$, 
Fig.~\ref{fig.Explosive_fractal} shows explosive growth at high persistence 
not only for $\beta = 1$ but also for $\beta = 2/3$, confirming the reduced
threshold. Finally, the prediction that the algebraic growth exponents in 
the diffusive regime are lower than in the compact case due to the influence 
of the fractal dimension is confirmed in 
Fig.~\ref{fig.DiffusiveAlgebraic_2d_fractal}, where results for algebraic 
scaling in the diffusive regime for $d = 2$ are shown.

\section{Conclusions}
\label{sec.conc}
In conclusion, we have investigated the scaling of cluster size with time for 
active particles in a solvent that irreversibly coagulate on collision by 
using theory and simulation. The scaling laws heavily depend on the scaling 
exponent $\beta$ of the total propulsion force of clusters. We identify four 
main scenarios for the total propulsion force scaling. If all particles in a 
cluster are aligned and able to contribute, the fastest growth is possible. 
Completely uncorrelated directions of particles lead to a significantly weaker 
scaling. Furthermore, hydrodynamics, fuel scarcity or lack of a field gradient 
required for propulsion can lead to the situation that only particles on the 
surface of a cluster can contribute. These contributing particles can again be 
completely aligned or their directions can be completely uncorrelated. The 
scaling of the total propulsion force is then further modified by the details 
of the propulsion mechanism and the thereby implied scaling exponent $\gamma$ 
of the single particle contribution force with the size of the cluster. 

Another crucial ingredient is the persistence length of cluster trajectories. 
In the diffusive regime, the persistence length is much smaller than the 
extension of clusters. Clusters explore the system volume and encounter each 
other on a diffusive time scale. More efficient growth occurs in the ballistic 
regime applying for a persistence length much larger than the cluster 
extension, where clusters sweep through the system volume on their 
semi--ballistic trajectories. The ballistic regime should be more relevant for 
active particle clusters since the persistence length of trajectories of 
active particles is usually rather large and tends to increase with aggregate 
size. 

We have verified these predictions in a simulation of compact clusters 
modelled as droplets that merge on contact in $d = 2,3$ dimensions. The 
simulation data shows good agreement with the model scaling laws and gives the 
correct algebraic growth exponents or exponential growth corresponding to the 
various values of $\beta$ in both regimes. 

Additionally, we extended our model to fractal clusters which 
show a different growth behaviour due to increased drag (hampering growth) 
and increased collision cross--section (enhancing growth). In the ballistic 
regime, the increased collision cross--section dominates the increased drag, 
leading to faster growth. However, in the diffusive regime, the increased drag 
dominates, resulting in a comparatively slower growth.

Given a sufficiently strong attraction between particles leading to 
irreversible coagulation, our findings are verifiable in experiments 
\cite{Theurkauff2012_PRL,Palacci2013_Science,Baraban2013_ACSNano}. Usually it 
is attempted to avoid attractions like van--der--Waals attraction appearing in 
metal--capped active particles. However, by intentionally enhancing the 
attraction to a level where particles cluster irreversibly the prerequisites 
of our theory can be met.
\acknowledgments
We thank Kurt Binder, Thomas Speck, Akira Onuki and Hajime Tanaka for helpful 
discussions. Financial support from the ERC Advanced Grant INTERCOCOS (Grant 
No. 267499) and the newly founded DFG Science Priority Program SPP 1726 is 
gratefully acknowledged.
\bibliography{references}
\end{document}